# Magneto-optical properties of textured La$_{2/3}$Sr$_{1/3}$MnO$_3$ thin films integrated on silicon via a Ca$_2$Nb$_3$O$_{10}$ nanosheet layer

Tomáš Maleček[1,2], Guillaume Agnus[2], Thomas Maroutian[2], Lukáš Horák[1], Petr Machovec[1], Valérie Demange[3], Aleš Melzer[1], Jan Prokleška[1], Philippe Lecoeur[2], Martin Veis[1]

1. Charles University, Faculty of Mathematics and Physics, Prague, Czech Republic
2. Université Paris-Saclay, CNRS, Centre de Nanosciences et de Nanotechnologies (C2N), 10 Boulevard Thomas Gobert, 91120 Palaiseau, France
3. Univ Rennes, CNRS, ISCR – UMR 6226, F-35000 Rennes, France

We demonstrate the possibility of growing textured La$_{2/3}$Sr$_{1/3}$MnO$_3$ (LSMO) thin films on silicon substrates with magneto-optical and optical properties comparable to high-quality epitaxial layers grown on bulk SrTiO$_3$ (STO). The pulsed laser deposition growth of LSMO is achieved by a two-dimensional nanosheet (NS) seed layer of Ca$_2$Nb$_3$O$_{10}$ (CNO) inducing epitaxial stabilization of LSMO films. The resulting layers possess a higher Curie temperature and a lower overall magnetization than samples of LSMO on STO. Spectra of the full permittivity tensor were calculated from optical and magneto-optical measurements. Spectral dependencies of both the diagonal and off-diagonal elements share many similarities between the LSMO/NS/Si and LSMO/STO samples. These similarities indicate comparable electronic structures of the layers and demonstrate comparable optical quality of textured LSMO on NS/Si and epitaxial LSMO on STO.

# Introduction

Hole-doped perovskite manganite La$_{1-x}$Sr$_x$MnO$_3$ has been at the centre of attention since the discovery of colossal magnetoresistance (CMR) in La-Ca-Mn-O films [1]. It contains trivalent and tetravalent manganese which leads to a double exchange mechanism (DE) responsible for unusual correlation between structural, magnetic, transport, and optical properties. It was found that the interplay between electron-phonon coupling arising from Jahn-Teller effects and DE influences essential physical properties of manganites [2]. When properly doped by Sr (x=1/3) the La$_{2/3}$Sr$_{1/3}$MnO$_3$ (LSMO) layers become metallic and ferromagnetic at room temperature (T$_C$ ~ 370 K in bulk [3]) with a high degree of spin polarization [4]. This makes LSMO an auspicious candidate for spintronic, optoelectronic, or magneto-optical applications. However, properties of LSMO thin films, especially magnetic properties, are strongly dependent on strain induced by the substrate [5-7]. The Mn-O bond length and angle strongly influence DE inducing notable shifts in T$_C$ [8]. Moreover, it takes several atomic layers to fully establish magnetic order in the LSMO film resulting in magnetically dead layers at the LSMO/substrate interface [9]. These phenomena influence the optical and magneto-optical responses [5] since they manifest themselves as small perturbances in the band structure of the material.

LSMO thin films are commonly deposited on nearly lattice-matched substrates such as SrTiO$_3$ (STO) or (LaAlO$_3$)$_{0.3}$(Sr$_2$AlTaO$_6$)$_{0.7}$ (LSAT) to achieve the highest possible crystalline quality and optimal physical properties [5, 10, 11]. Nevertheless, there were also attempts to grow LSMO thin films on lattice-mismatched substrates, such as Si. Such films deposited by magnetron sputtering have been reported as amorphous [12] or polycrystalline [13] and their magneto-optical properties were significantly suppressed compared to those of epitaxial layers. An improvement has been observed with the use of a buffer layer. Materials such as SrTiO$_3$ (STO) [14-16], CaTiO$_3$ [14], or a CeO$_2$/ YSZ (yttria-stabilized zirconia) multilayer [17, 18] have been investigated as possible buffer layers resulting in epitaxial growth of LSMO. However, the growth process relies upon advanced and expensive deposition techniques that are not well scalable for production due to their complexity.

Another approach to perovskite deposition onto lattice-mismatched substrates utilizes a nanosheet (NS) seed layer [19-21]. This approach has been recently used to demonstrate the growth of textured LSMO films on glass substrates with a Ca$_2$Nb$_3$O$_{10}$ (CNO) NS seed layer. The resulting films exhibited magnetic properties close to epitaxial layers on STO [22]. Better insight into the electronic structure of LSMO layers deposited on the CNO NS seed layer could be gained through their optical and magneto-optical properties, which have not been studied yet.

In this letter, we report on the growth of high-quality (001)-textured LSMO films on Si using the CNO NS seed layer. We systematically studied their magnetic, optical, and magneto-optical properties and compared them to those of an epitaxial LSMO layer on (001) STO. The results indicate a high optical quality of the LSMO films on Si substrates.

# Experimental details

Colloidal solution of Ca$_2$Nb$_3$O$_{10}$ nanosheets has been obtained by exfoliation of the parent layered KCa$_2$Nb$_3$O$_{10}$ phase as described in [22]. The nanosheets were deposited on Si, with a thin native oxide layer on top, using a drop-casting method described in [23]. For this purpose, a mixture of 120 µL of nanosheet colloidal solution, 15 mL of ultra-pure water, and 600 µL of absolute ethanol was prepared and drop-casted on the substrate heated at 100°C on a hot plate.

LSMO thin films have been grown from a stoichiometric ceramic target by pulsed laser deposition (PLD) using a KrF laser operating at λ = 248 nm on (001) oriented STO and NS/Si substrates. The substrate was kept at about 650 °C. Oxygen was used as a background gas during the deposition. The pressure was maintained at 110 mTorr. Post-deposition cooling to room temperature was done at 75 Torr.

Crystalline properties, surface morphology, and magnetic properties of the samples were examined using X-ray diffraction (XRD) and reflectometry (XRR), atomic force microscopy (AFM), and vibrating-sample magnetometry (VSM). Symmetric XRD measurements utilized a PANanalytical X'Pert PRO diffractometer in parallel beam configuration. XRR and asymmetric XRD experiments were carried out with a Rigaku Smartlab diffractometer equipped with a 9kW rotating Cu anode. The measurements were performed in a parallel beam geometry using a parabolic X-ray mirror (Cu K-alpha radiation). The AFM images were captured by a Bruker Innova AFM microscope and the magnetic properties were measured by a Physical Property Measurement System from Quantum Design equipped with a VSM option and a 9 T superconducting coil.

The optical response of the samples was measured by a Woollam RC2 ellipsometer in the spectral range from 0.74 eV to 6.4 eV. Angles of incidence ranged from 55° to 75° in 5° increments. The magneto-optical response was acquired by a custom-built magneto-optical Kerr effect (MOKE) spectrometer with precision below 1 mdeg. Its spectral range was from 1.4 eV to 5 eV at an applied magnetic field of 1 T in polar geometry (out-of-plane). The permittivity tensor of the sample in such geometry takes the following form

$$\boldsymbol{\varepsilon} = \begin{pmatrix} \varepsilon_1 & i\varepsilon_2 & 0 \\ -i\varepsilon_2 & \varepsilon_1 & 0 \\ 0 & 0 & \varepsilon_1 \end{pmatrix} \quad (1)$$

The spectral dependence of its elements describes the optical and magneto-optical response and is directly related to the band structure of the material. Therefore, their knowledge is crucial for the description of small strain-related changes in the orbital ordering of LSMO due to the use of the NS buffer layer.

## Results and Discussion

**Structural properties.** The morphology of the NS/Si substrate and the LSMO film as examined by AFM is shown in Figure 1. Surface coverage of the Si substrate by the NS layer was estimated to be around 90 %. About 10 % of the surface is covered by a second layer of NS. The size of the NS as an equiaxed shape ranges from 40 nm to 500 nm. The thin layer of LSMO follows the morphology of the NS, with the edges of the LSMO-covered NS discernible on the AFM images. The root mean square (RMS) roughness of the LSMO/NS/Si sample is 0.74 nm.

The symmetric $\theta$-$2\theta$ XRD pattern of the LSMO/NS/Si sample (Figure 1(c)) revealed the presence of the (001) oriented LSMO phase. More detailed symmetric and asymmetric scans (not shown here) revealed the pseudocubic out-of-plane parameter to be $c$ = 0.3860 nm which is close to the bulk value of LSMO (c = 0.3876 nm [24]). The in-plane parameters are a = b = 0.3885 nm. The XRD patterns confirmed the growth of LSMO on NS with the out-of-plane (001) fiber texture. The crystallites are randomly rotated in-plane around the out-of-plane [001] orientation. Furthermore, from the XRR (Figure S1) the thickness of the LSMO film was fitted as $t$ = 40.3 nm with an additional 5.4 nm interface layer between the LSMO and the natural 2.3 nm thick $SiO_2$ oxide atop Si. Since the NS do not cover the silicon substrate fully the interface layer comprises a mixture of the 1.4 nm thick NS and LSMO with voids. The 40 nm thick LSMO layer above this interface layer is compact and homogenous without voids.

The STO substrate exhibited surface atomic steps equivalent to a 0.12° miscut (see Figure S2). The LSMO film on STO follows the morphology and atomic steps of the substrate demonstrating epitaxial growth. The RMS roughness of the LSMO/STO samples was 0.54 nm, which is close to the NS sample and demonstrates the high surface quality of the layers.

The out-of-plane lattice parameter of the LSMO/STO sample $c$ = 0.3844 nm and film thickness $t$ = 49.6 nm were determined from the $\theta$-$2\theta$ XRD pattern by the classic interference formula (see Figure S2(c)). While the LSMO thin film on the NS buffer layer is almost negligibly strained, the LSMO layer on the STO substrate is under in-plane tensile strain as expected [5].

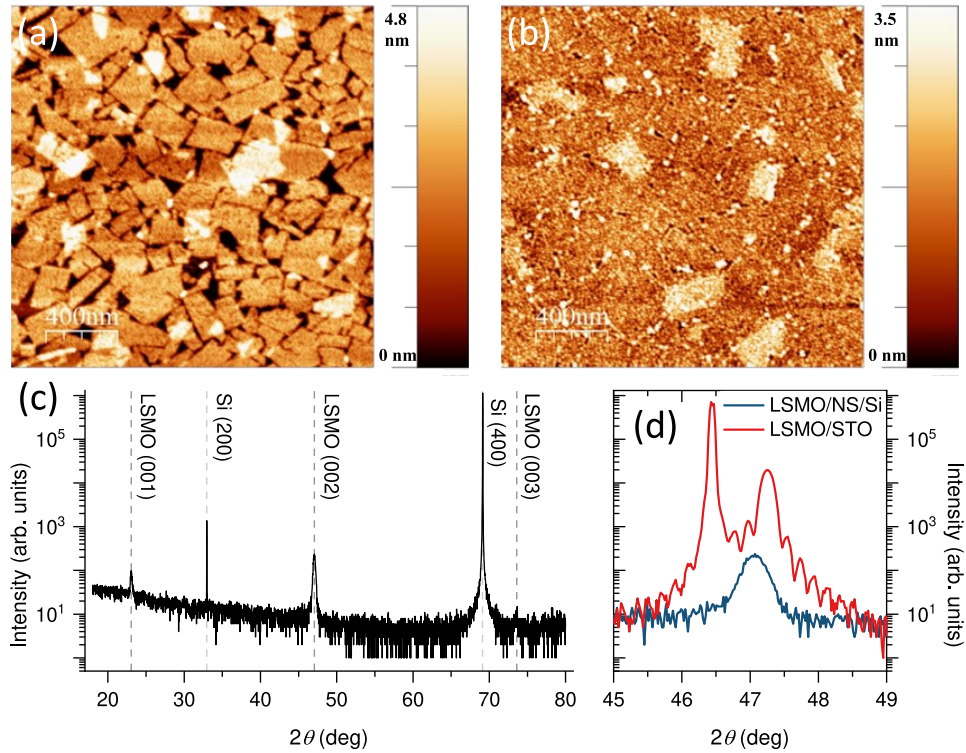

*Figure 1 AFM images of (a) the NS/Si substrate and (b) the 45 nm thin film of LSMO atop NS/Si. A symmetric θ-2θ XRD pattern of the LSMO/NS/Si sample is shown in (c). XRD confirms the out-of-plane (001) orientation of LSMO. Figure (d) compares the (002) diffraction peak of LSMO grown on NS/Si (blue) with LSMO grown on STO (red). Laue oscillations are observable around the (002) diffraction peak of LSMO on STO along with the (002) diffraction peak of STO.*

**Magnetic properties.** The temperature dependence of saturated magnetization is shown in Figure 2. A magnetic field of 0.2 T has been applied in-plane to the samples whilst cooling down from 400 K to 3 K. Magnetic hysteresis loops are shown in Figure S3. The saturated magnetization of the LSMO/NS/Si sample reaches only 70 % of that of LSMO on STO at 5 K. On the other hand, the LSMO/NS/Si sample possesses a higher Curie temperature, $T_C$ = 362 K, than LSMO/STO, $T_C$ = 352 K, due to its strain relaxation (see Figure S4). The epitaxial strain has been shown to deteriorate the magnetic properties of thin LSMO films [6] and has been linked to lower $T_C$ in films than in bulk. The value of low-temperature saturation magnetization of LSMO/NS/Si agrees with LSMO films deposited on NS/glass by Boileau *et al.* [22]. A decrease in magnetization has been observed in polycrystalline LSMO as a result of an increased number of magnetically disordered states at the grain boundaries [25]. The magnetization also depends on the grain size of the substrate in LSMO films grown on lattice-matched polycrystalline substrates [26].

The coercive field, $H_C$ = 23 Oe, at room temperature (see Figure S3) is larger for the LSMO/NS/Si sample compared to the epitaxial film of LSMO on STO with $H_C$ = 3 Oe. The increase in $H_C$ can be tied to studies of polycrystalline LSMO films [27] where each grain switches independently leading to a higher macroscopic coercivity. Since the LSMO film on top of the NS exhibits columnar growth, in-plane magnetization switching resembles a polycrystalline film.

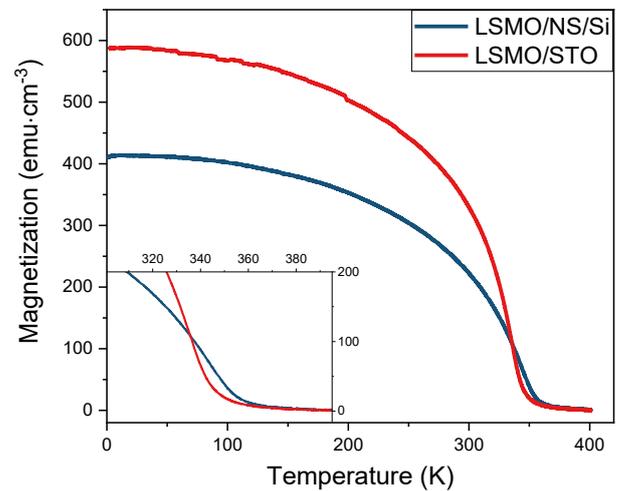

*Figure 2 Magnetization as a function of temperature for an applied in-plane field of H = 0.2 T.*

**Optical properties.** Experimental ellipsometric data (see Figure S5) have been fitted by CompleteEASE software to deduce the spectral dependence of optical parameters (diagonal elements of the permittivity tensor) of the investigated films. For the description of all spectral dependencies of $\varepsilon_1$ a B-spline function [28] was used. The NS-covered Si substrate was measured separately prior to the LSMO sample and has been modelled as a semi-infinite Si substrate (optical properties taken from [29]) covered with a 2.3 nm thick native oxide layer (optical properties taken from [29]) and a 1.4 nm thick NS layer. This allowed for the characterization of the optical response of the NS layer, which was later utilized in subsequent fitting of the LSMO sample. The LSMO layer was parametrized similarly by a B-Spline function considering a model structure of LSMO/NS/Si with a fitted LSMO thickness of 46.5 nm. This is in good agreement with the XRR measurements since the optical model does not consider the interface LSMO layer as independent. The model structure of the LSMO/STO sample consisted of a semi-infinite substrate with a thin film on top. The optical properties of the STO substrate were taken from bare substrate measurements. The spectral dependencies of the diagonal element $\varepsilon_1$ of the permittivity tensor for LSMO/NS/Si and LSMO/STO are displayed in Figure 3. The spectral shape of $\varepsilon_1$ of the LSMO film on the STO substrate follows already reported data [5, 30-32]. The spectral dependence of $\varepsilon_1$ of the LSMO/NS/Si sample doesn't differ much from LSMO/STO. The maximum of the peak in the imaginary part of $\varepsilon_1$ around 4 eV is lower and slightly shifted in energy. This spectral structure is usually ascribed to an O 2p → Mn 3d $t_{2g}$ ↓ transition [13, 33-35]; however, two peaks have been shown to be a better fit for this optical feature [5]. It has also been shown that epitaxial strain induces energy shifts and amplitude modulation of this spectral feature [5]. Larger deviance in $\varepsilon_1$ spectra among the samples is notable in the infrared part of the spectrum which corresponds to the O 2p → Mn 3d $e_g$ ↑ transition [13, 35], suggesting a minor energy shift of the involved Mn orbital. The maximum of the corresponding spectral structure, which is out of the measured range, is probably shifted to lower energy in the LSMO/NS/Si sample. Additional infrared measurements are necessary to validate this claim. The infrared part of the LSMO/NS/Si dependence follows a similar trend as that of amorphous LSMO layers grown on Si (111) by Monecke *et al.* [12]. However, the spectral features of LSMO amorphous layers are broader and the absolute values of Re{$\varepsilon_1$} of the amorphous layers are systematically higher whereas the absolute values of Im{$\varepsilon_1$} are systematically lower. We can therefore conclude that the minor differences in optical properties originate from strain relaxation in LSMO/NS affecting mainly O 2p → Mn 3d $e_g$ ↑ transition.

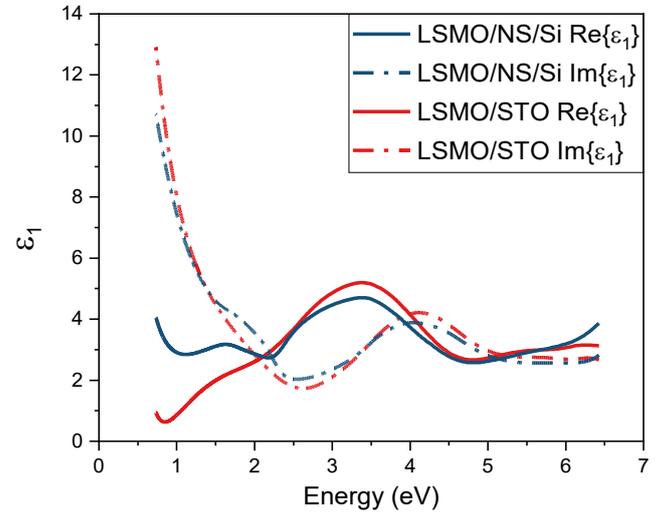

*Figure 3 Relative permittivity spectra of LSMO/NS/Si (blue) and LSMO/STO (red).*

**Magneto-optical properties.** MOKE spectra have been measured in polar configuration (see Figure S6). The off-diagonal element $\varepsilon_2$ of the permittivity tensor has been calculated from MOKE and $\varepsilon_1$ using Yeh's 4x4 matrix formalism [36]. The spectra of $\varepsilon_2$ are shown in Figure 4. The LSMO/STO sample exhibits spectral behaviour already reported in literature [5, 31, 34], with a dominating diamagnetic transition at around 3.6 eV. The LSMO/NS/Si sample has a similar spectral shape and a lower amplitude (accordingly to lower saturation magnetization). However, the amplitude of the LSMO/NS/Si film is still higher than that of amorphous LSMO films prepared by magnetron sputtering by Monecke *et al.* [12]. The main transition at around 3.6 eV is shifted to lower energies compared to the LSMO/STO sample. It has been reported that the shift of the main transition to lower energies originates in lower epitaxial strain in the thin film of LSMO [5].

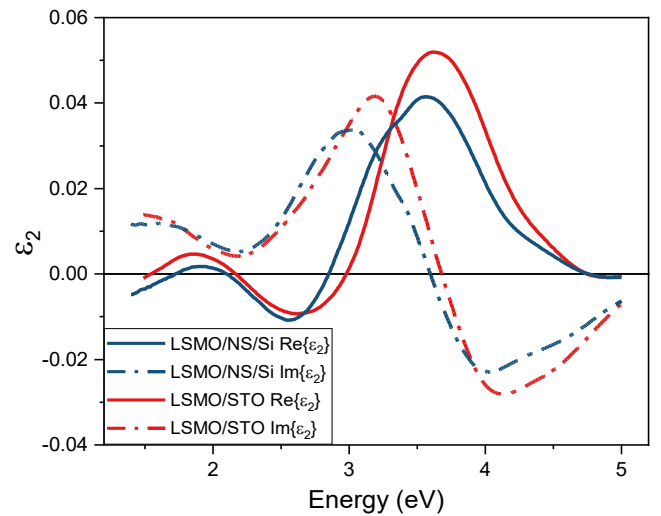

*Figure 4 Spectra of the off-diagonal element of the permittivity tensor of LSMO/NS/Si (blue) and LSMO/STO (red) in polar geometry.*

# Conclusions

Physical properties of an LSMO thin film integrated on silicon via a nanosheet seed layer have been studied. AFM and XRD measurements confirmed the (001)-texture of the LSMO film. Magnetic measurements revealed the increase of Curie temperature by 10 K compared to a reference epitaxial LSMO film on STO. The rise in Curie temperature has been explained by strain relaxation of the textured LSMO layer. A 30 % decrease in saturated magnetization at 5 K has been observed with respect to the epitaxial film. This has been linked to magnetically disordered states at grain boundaries and in the areas between particular NS. Spectral dependence of the full permittivity tensor has been deduced from optical and magneto-optical measurements and found to be similar for both NS/Si and STO samples. Comparable optical and magneto-optical response of both samples indicate their similar electronic structures. A deviation in the spectral shape of $\varepsilon_1$ was observed in the infrared region suggesting a minor energy shift of Mn 3d $e_g \uparrow$ orbital due to lower strain. The spectral dependence of $\varepsilon_2$ of LSMO/NS/Si exhibits a lower amplitude than that of LSMO/STO. This agrees with magnetization measurements. The main spectral structure in the $\varepsilon_2$ spectra of the LSMO/NS/Si is shifted to lower energies compared to the LSMO/STO maximum. This is the result of the strain relaxation in the LSMO layer on silicon.

Finally, our results confirm the growth of an LSMO thin film on Si substrate with optical and magneto-optical properties close to epitaxial films promising the possibility to growth LSMO on various non-compatible substrates. This is important for potential spintronic applications.

# Supplementary material

Refer to the supplementary material for additional information and supporting experimental analyses of the samples. The supplementary material includes XRR of the LSMO/NS/Si sample (Figure S1), structural characterization of the LSMO/STO sample (Figure S2), magnetic hysteresis loops (Figure S3), method used for determining the Curie temperatures of the samples (Figure S4), the measured ellipsometric angles (Figure S5), and MOKE spectra (Figure S6).

# Acknowledgments

The study was supported by the Charles University, project GA UK No. 363321.

This work was supported by the Charles University grant SVV–2023–260720.

Magnetization experiments were performed in MGML (mgml.eu), which is supported within the program of Czech Research Infrastructures (project no. LM2023065).

This work has received support from the French national network RENATECH.

We thank the CRISMAT laboratory (Caen, France) for providing the LSMO ceramic target.

# Author Declarations

## Conflict of Interest

The authors have no conflicts to disclose.

## Author Contributions

**Tomáš Maleček:** Conceptualization (supporting); Data Curation (lead); Formal Analysis (equal); Funding Acquisition (lead); Investigation (equal); Project Administration (equal); Validation (equal); Visualization (lead); Writing/Original Draft Preparation (lead); Writing/Review & Editing (supporting). **Guillaume agnus:** Conceptualization (lead); Formal Analysis (supporting); Project Administration (equal); Resources (equal); Supervision (equal); Validation (equal); Writing/Original Draft Preparation (supporting); Writing/Review & Editing (equal). **Thomas Maroutian:** Formal Analysis (supporting); Investigation (supporting); Methodology (equal); Project Administration (supporting); Resources (equal); Supervision (equal). **Lukáš Horák:** Data Curation (supporting); Formal Analysis (equal); Investigation (equal); Methodology (equal); Resources (equal); Supervision (supporting); Validation (equal); Visualization (supporting). **Petr Machovec:** Data Curation (supporting); Formal Analysis (equal); Investigation (equal); Methodology (equal); Validation (equal); Visualization (supporting); Writing/Review & Editing (supporting). **Valérie Demange:** Conceptualization (supporting); Methodology (equal); Resources (equal); Writing/Original Draft Preparation (supporting). **Aleš Melzer:** Investigation (supporting); Validation (supporting). **Jan Prokleška:** Formal Analysis (supporting); Methodology (supporting); Resources (equal); Supervision (supporting). **Philippe Lecoeur:** Methodology (supporting); Resources (equal). **Martin Veis:** Conceptualization (supporting); Data Curation (supporting); Formal Analysis (equal); Funding Acquisition (supporting); Methodology (equal); Project Administration (equal); Resources (equal); Supervision (equal); Validation (equal); Writing/Original Draft Preparation (supporting); Writing/Review & Editing (equal).

# Data availability

The data that support the findings of this study are available from the corresponding authors upon reasonable request.

# Supplementary Information

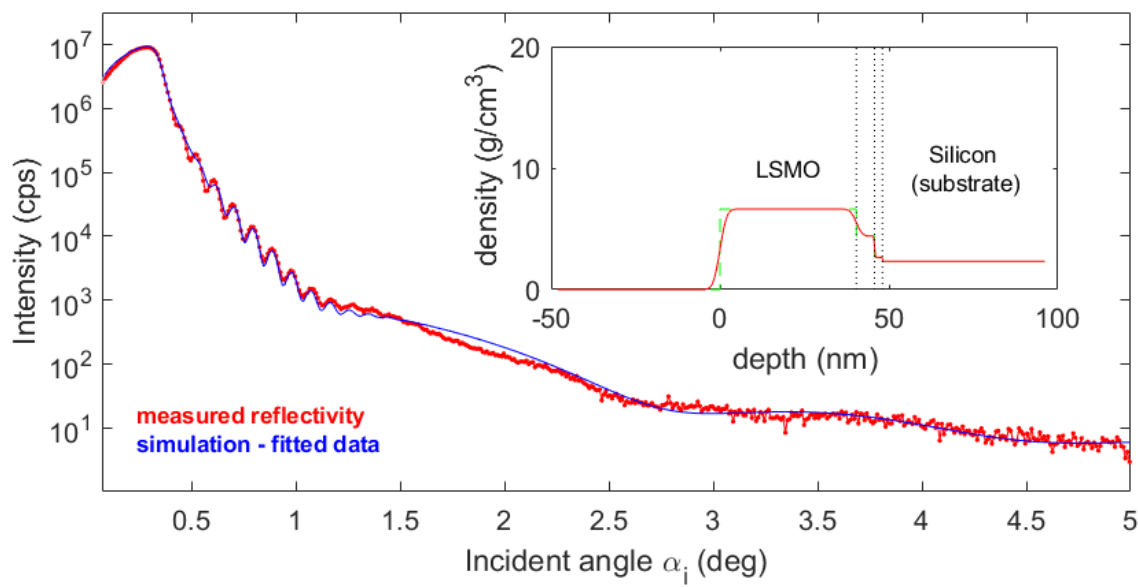

*Figure 1* X-ray reflectivity curve measured up to incidence angle of 5 deg (red) fitted by the simulation (blue) based on the dynamical theory of diffraction assuming homogeneous layers with rough interfaces. The high-frequency oscillations are caused by the finite thickness of the LSMO layer, while low-frequency thickness fringes correspond to the presence of the low-density transition layer between the compact LSMO layer and the substrate. The determined density depth-profile of the sample is plotted in the inset.

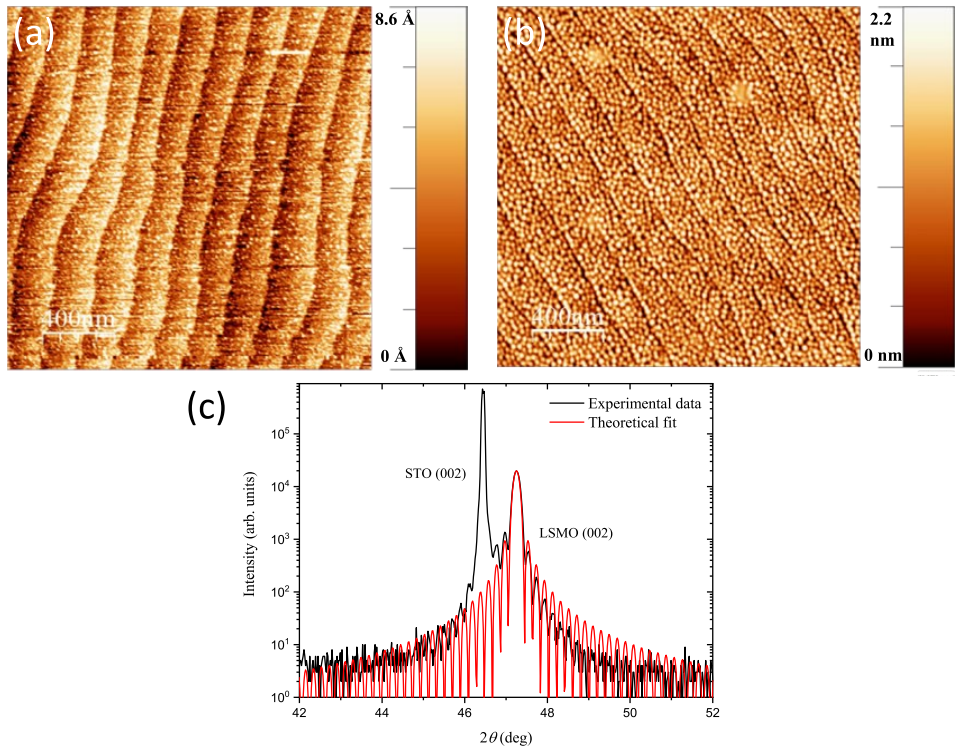

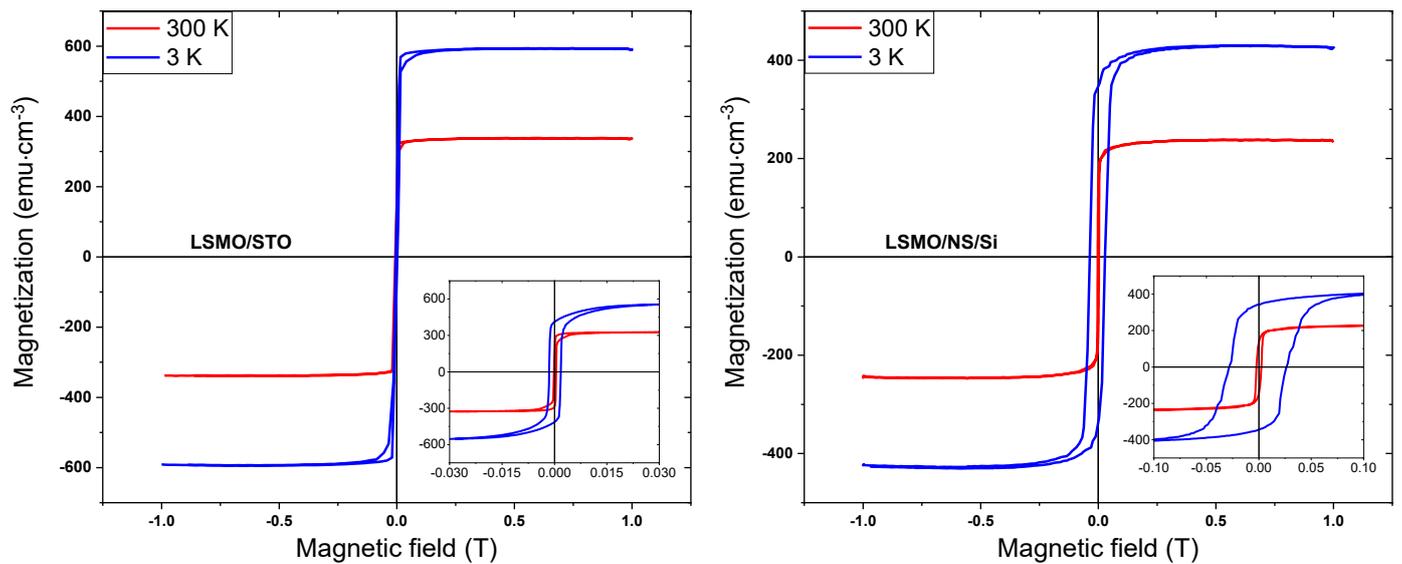

*Figure 2* AFM images of (a) the STO substrate and (b) the thin film of LSMO atop STO. Atomic steps of the substrate correspond to a miscut angle of 0.12°. A symmetric θ-2θ XRD scan of the LSMO/STO sample is shown in subfigure (c) along with the theoretical fit of the Laue oscillations corresponding to a thickness of 49.6 nm.

*Figure 3* In-plane hysteresis loop measurements of an LSMO/STO sample (left) and an LSMO/NS/Si sample (right) at 300 K (red) and 3 K (blue).

Curie temperature has been determined using the two-tangent method [37]; however, it was applied to the M(T) relation in the log-log scale to better reflect the behaviour of magnetization below $T_C$ (see Figure S4).

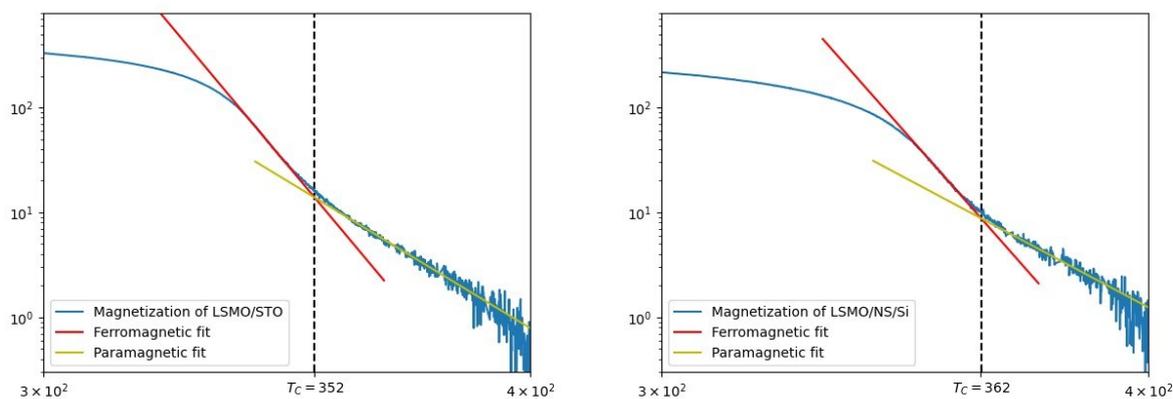

*Figure 4 Magnetization as a function of temperature, in a log-log scale, of an LSMO/STO sample (left) and an LSMO/NS/Si sample (right). Fitting both sides of the transition by a linear function in a logarithmic scale yields different gradients. The Curie temperature has been determined as the temperature of the crossing point of these linear functions.*

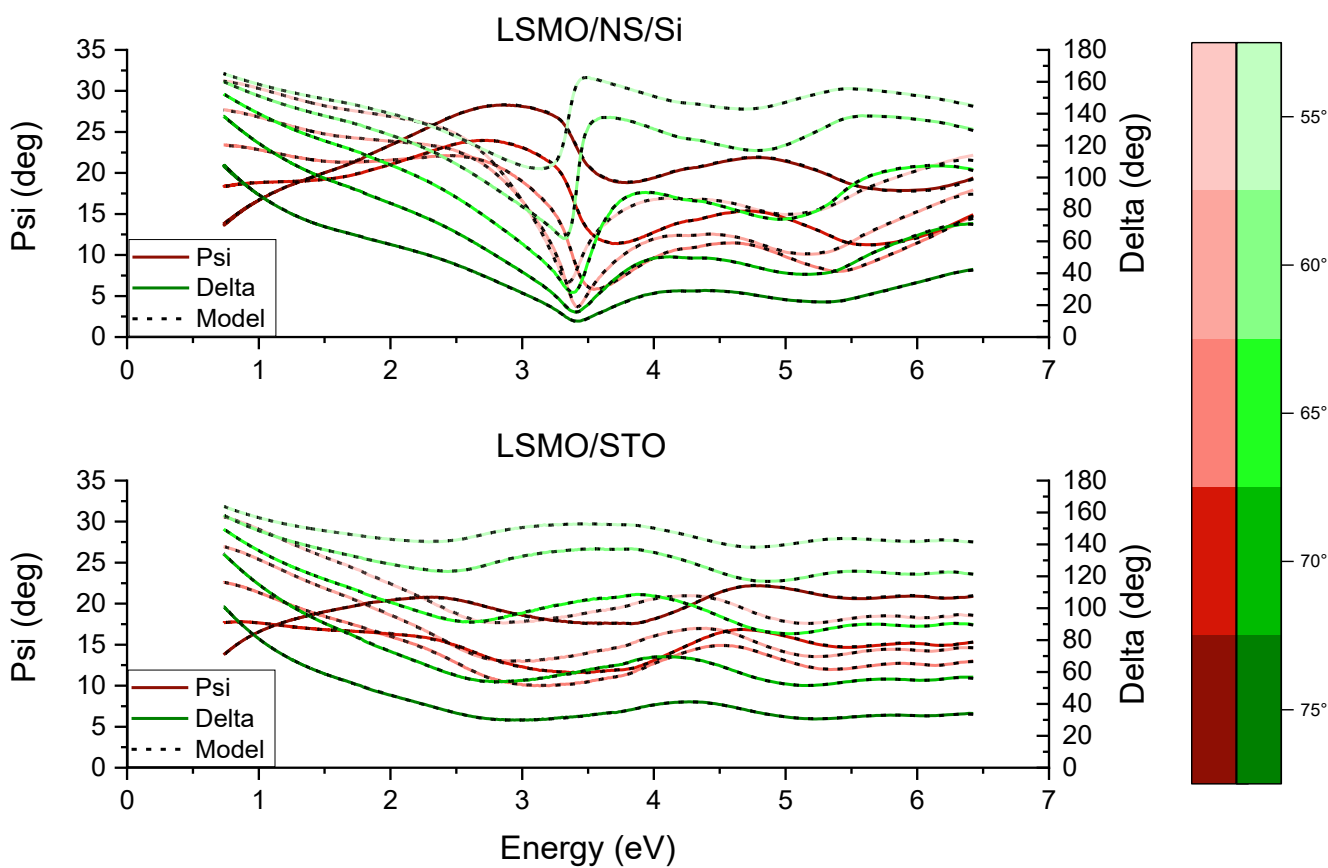

*Figure 5 Ellipsometric angles Psi and Delta as a function of energy for the LSMO/NS/Si sample (top) and LSMO/STO sample (bottom). Line colour corresponds to the angle of incidence.*

MOKE spectra have been measured in polar configuration (see Figure S6) to determine the off-diagonal element of the permittivity tensor of LSMO. Spectra of both samples exhibit a notable minimum in Kerr rotation at around 3.3 eV and a smaller global maximum at around 2.6 eV. The spectral dependence of Kerr ellipticity shows a corresponding 'S' shaped spectral structure centered around 3.3 eV. Spectral features of the LSMO/NS/Si sample are narrower than those of LSMO/STO and exhibit a considerably larger MOKE amplitude. The larger amplitude is the result of optical effects. The MOKE spectra of the LSMO/NS/Si sample resemble spectra of LSMO deposited on amorphous Si and annealed at 800 °C by Monecke *et al.* [12]. The spectra of LSMO/STO are in agreement with the literature [5].

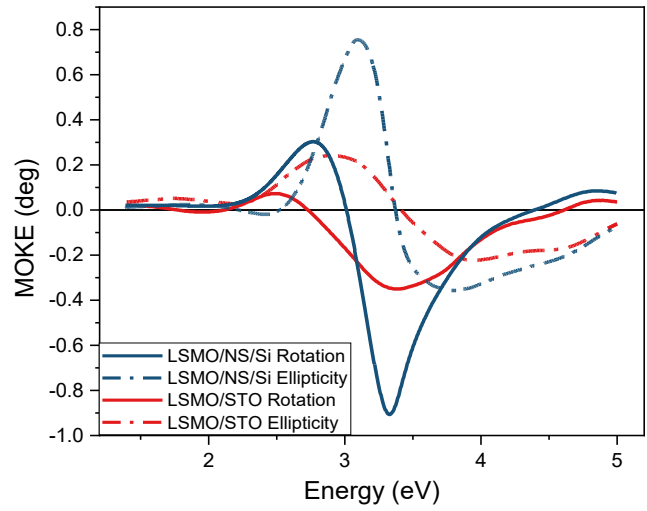

***Figure 6*** *Spectra of Kerr rotation and Kerr ellipticity of the LSMO/NS/Si sample (blue) and the LSMO/STO sample (red).*